\documentclass[twoside]{ilcws08}
\usepackage[latin1]{inputenc}
\usepackage[dvips]{graphicx,epsfig,color}
\usepackage{wrapfig,rotating}
\usepackage{amssymb,amsmath,array}

\begin{document}
\title{GEM studies for LCTPC}
\author{Lea Hallermann$^{1,2}$ and Jeannine Beck$^{2,3}$
\vspace{.3cm}\\
1- Universit\"{a}t Hamburg - Institut f\"{u}r Experimentalphysik\\
Luruper Chaussee 149, 22761 Hamburg - Germany
\vspace{.1cm}\\
2- DESY, Deutsches Elektronen-Synchrotron in der Helmholtz-Gemeinschaft\\
Notkestrasse 85, 22607 Hamburg - Germany
\vspace{.1cm}\\
3- Martin-Luther-Universit\"{a}t Halle-Wittenberg - Institut f\"{u}r Physik\\
Friedemann-Bach-Platz 6, 06108 Halle - Germany 
}

\maketitle

\begin{abstract}
The LCTPC collaboration is developing a time projection chamber (TPC) for a future linear collider.
In the ILD concept for an ILC detector a TPC is foreseen as main tracking device. 
In order to get measurable signals on the anode, the primary electrons, produced by charged particles traversing the TPC, have to be multiplied. For this purpose gas electron multipliers (GEMs)
can be used. Studies of these GEMs are performed at the DESY FLC group and presented in the 
following.
\end{abstract}

\section{Comparison of Different GEM Types}
GEMs are produced by a number of different manufacturers. They differ in material, hole shape and hole pitch.
For comparative studies of GEMs five types have been used. All important parameters are listed in table \ref{tab:gems}.
The CERN GEM with $140~\mu$m hole pitch is used as standard in this study.
Another type of CERN GEM with a larger pitch of $225~\mu$m was available, as well as GEMs from the TechEtch company \cite{te} in the US. They are similar to standard CERN GEMs.
Furthermore two different thick GEMs from the Japanese company SciEnergy \cite{se} have been tested. They use a liquid crystal polymer (LCP) instead of polyimide as substrate and possess cylindrical instead of double conical holes. 
One sample has a $50~\mathrm{\mu m}$ thick LCP insulator, the other $100~\mathrm{\mu m}$. Liquid crystal polymer shows less water absorption than polyimide and a very small thermal expansion coefficient.\\
All measurements were performed with a small TPC prototype. It has a drift distance of $20$~mm, a double GEM stack operated by a voltage divider 
and an unsegmented anode. The gas mixture is composed of $93\%$ argon, $5\%$ methan and $2\%$ CO$_{2}$. For the drift field $250$~V/cm are applied and the transfer and induction fields are set to $1$ kV/cm. An iron-55 gamma source mounted at the cathode is used as signal source. The measured charge is preamplified with a charge sensitive amplifier and read out with a charge to digital converter.\\

\begin{table}
\begin{tabular}{c c r r c c} 
\hline 
 \multicolumn{1}{c}{manufacturer} & \multicolumn{1}{c}{substrate} & \multicolumn{1}{c}{thickness} & \multicolumn{1}{c}{pitch} & \multicolumn{1}{c}{etching} & \multicolumn{1}{c}{hole shape}\\
\hline
 CERN GDD group & polyimide & $50\mu$m & $140\mu$m & chemical & double conical \\ 
 CERN GDD group & polyimide & $50\mu$m & $225\mu$m & chemical & double conical \\ 
 TechEtch, USA & Kapton\textsuperscript{\textregistered} & $50\mu$m & $140\mu$m & chemical & double conical\\
 SciEnergy, Japan & LCP & $50\mu$m & $140\mu$m & Laser/plasma & cylindrical\\
 SciEnergy, Japan & LCP & $100\mu$m & $140\mu$m & Laser/plasma & cylindrical\\
 \hline
 \end{tabular}
\caption{GEM types and their parameters.}
\label{tab:gems}
\end{table}

\subsection{Gain}
\begin{wrapfigure}{r}{0.5\columnwidth}
\centerline{\includegraphics[width=0.45\columnwidth,clip=]{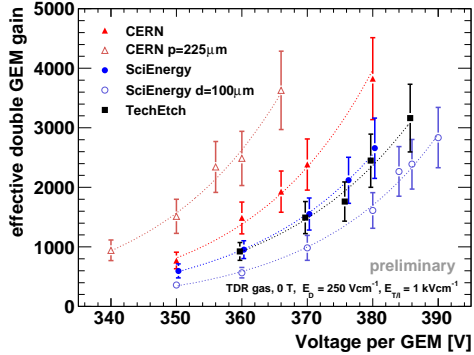}}
\caption{Effective double GEM gain of the five different types over voltage per GEM. 
The lower GEM is always a CERN standard GEM. Exponential functions are fitted to the data.
}\label{fig:gain}
\end{wrapfigure}

The gain of the setup is measured based on known properties of Iron-55. The results are compared to a gain parametrization published in \cite{para}. Good agreement is found.\\
Figure \ref{fig:gain} shows the measured gains for the different GEM types with exponential functions fitted to the data. Each measurement uses the same standard CERN GEM in the lower position, only the upper one was replaced. The CERN GEMs produce the highest gain. Increasing the pitch at a fixed voltage increases the gain. This is presumably due to the higher surface charge at the polyimide on the hole walls, due to the smaller number of holes. The larger surface charge causes a higher field in the hole center, resulting in a higher gain at the same voltage.\\
The reason for the lower gain of the TechEtch GEM is not yet clear. The only difference to the standard CERN GEMs is a different polyimide type used for the insulation layer.\\
The LCP based GEMs have at a fixed voltage a lower gain than the standard CERN GEMs. As expected the gain decreases if the substrate thickness is increased. In the case of the $50~\mathrm{\mu m}$ thick GEM similar values as for the TechEtch GEM are obtained. The thicker GEM has at the same voltage a smaller field inside the holes leading to a smaller gain.

\subsection{Energy Resolution}

\begin{wrapfigure}{r}{0.5\columnwidth}
\centerline{\includegraphics[width=0.45\columnwidth,clip=]{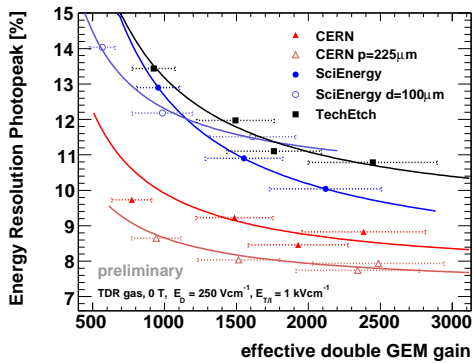}}
\caption{Energy resolution over effective double GEM gain. Data points are fitted with an hyperbolic function.}\label{fig:er}
\end{wrapfigure}

To achieve a better understanding of the quality of the GEMs the energy resolution was analyzed in addition to the gain. The energy resolution is obtained by an analysis of the Iron-55 spectra using the following formula: $ \sigma/E=\sigma_{p}/\mu_{p}$, where $\sigma_{p}$ denotes the Gaussian width of the photo peak in the spectra and $\mu_{p}$ the peak position.\\
Figure \ref{fig:er} shows the energy resolution as a function of the gain. The data points are fitted with an hyperbolic function. Resolutions between 8\% and 14\% were found. The CERN GEMs produce the best energy resolution. In order to explain the differences in particular between GEMs with double conical and cylindrical holes, electrostatic field simulations were performed \cite{jean}. They show that the gradient from the center of the hole to the walls is higher for cylindrical holes. That causes a higher energy spread for electrons amplified in GEMs with cylindrical holes with respect to the energy spread in GEMs with double conical holes.\\
The reason for the high values of energy resolution for the TechEtch GEMs has to be studied in more detail.

\section{Development of a GEM Support Structure}

\begin{wrapfigure}{r}{0.5\columnwidth}
\centerline{\includegraphics[width=0.45\columnwidth,clip=]{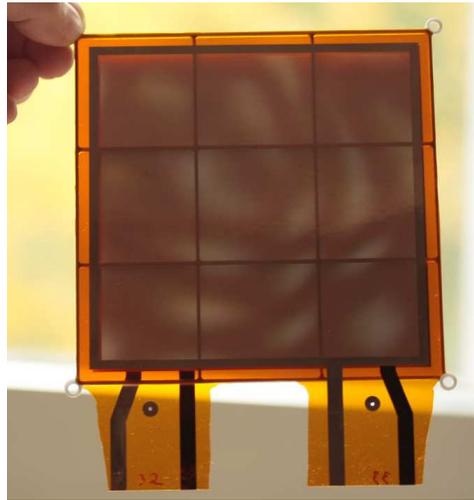}}
\caption{Photo of double GEM stack with grid glued in between the GEMs.}\label{fig:grid}
\end{wrapfigure}

A general problem of a GEM based TPC is the mounting of GEMs in readout modules. This mounting scheme should be lightweight, maintain enough tension in the GEM to ensure flatness, and have no dead zones. Traditionally external frames have been used, which however are not optimal for a large area readout system like the ILD TPC.

\subsection{Ceramic Grid}
A new approach has been to introduce a spacer grid between the GEMs outside the sensitive area. Great care has to be taken to control the dispensing of the glue to avoid high voltage problems caused by glue in the holes. Hence a commercial dispensing robot was used. The grids are manufactured from aluminum oxide ceramics. The bars are $1$~mm wide and $2$ mm high, while the size of the grid cell is $3.7 \times 3.7$ cm$^2$. At the moment the width of the bars is limited to $1$~mm, but studies are under way to reduce this to $0.5$ mm or even less. The glued structure, which can be seen in figure \ref{fig:grid}, is mechanically stable, lightweight and has only small dead zones.\\
First tests of these structures are very promising. The stack stands high voltage. Gain and energy resolution are comparable to the standard way of mounting.
In the future spatially resolved measurements are planned to study the impact of the ceramic bars on the overall performance in more detail.

\begin{footnotesize}

\end{footnotesize}

\end{document}